\documentstyle[12pt,aaspp]{article}

\begin{document}

\title{Simulations of Relativistic Extragalactic Jets}
\author{G. Comer Duncan\altaffilmark{1}}
\affil{Department of Physics and Astronomy, Bowling Green State University,
Bowling Green, OH 43403}
\author{Philip A. Hughes\altaffilmark{2}}
\affil{Astronomy Department, University of Michigan, Ann Arbor, MI 48109}
\altaffiltext{1}{E-mail: gcd@riemann.bgsu.edu}
\altaffiltext{2}{E-mail: hughes@astro.lsa.umich.edu}

\vskip 1.0cm
\centerline{Submitted to Astrophysical Journal Letters 7 June 1994}
\vskip 1.0cm

\begin{abstract}
 We describe a method for the numerical solution of the relativistic Euler
equations which we have found to be both robust and efficient, and which has
enabled us to simulate relativistic jets. The technique employs a solver of
the Godunov-type, with approximate solution of the local Riemann problems,
applied to laboratory frame variables. Lorentz transformations provide the
rest frame quantities needed for the estimation of wave speeds, etc. This is
applied within the framework of an adaptive mesh refinement algorithm,
allowing us to perform high-resolution, 2-D simulations with modest computing
resources. We present the results of nonrelativistic, and relativistic
($\gamma=5$ and $10$) runs, for adiabatic indices of $5/3$ and $4/3$. We find
the same gross morphology in all cases, but the relativistic runs exhibit
little instability and less well-defined structure internal to the jet: this
might explain the difference between (relatively slow) BL~Lacs and (faster)
QSOs. We find that the choice of adiabatic index makes a small but
discernible difference to the structure of the shocked jet and ambient
media.
\end{abstract}


\section{Introduction}

 There is now compelling evidence for the existence of relativistic flow
speeds in the jets associated with AGN (Cawthorne 1991).  Since the early
1980s numerical hydrodynamical simulations have been used to study such
systems, although most of this work has been restricted to the
nonrelativistic domain -- indeed, the difficulty of numerical hydrodynamics
in the relativistic domain has been discussed by Norman \& Winkler (1986).
Wilson (1987) and Bowman (1993) have explored steady relativistic jets, but
the interpretation of observations on the parsec and sub-parsec scale depend
critically on understanding the behavior of propagating structures. Although
time-dependent relativistic jet simulations have recently been published by
van Putten (1993) and Mart\'\i, M\"uller, \& Ib\'a\~nez (1994), this field
remains largely unexplored. However, the results of heavy-ion collisions may
be modeled using relativistic hydrodynamics (Schneider et al. 1993; Dean et
al. 1994), and a significant body of experience in solving the relativistic
Euler equations has accumulated in this field. In this paper we present the
first results from a code written to study extragalactic jets, employing an
Euler equation solver suggested by nucleon collision modeling, and applied
within the framework of an adaptive mesh refinement algorithm. In \S 2 we
describe the solver and in \S 3 the grid adaption technique; in \S 4 we
compare results for nonrelativistic and relativistic flows.

\section{Techniques}
The simulations are performed assuming axisymmetry, using as physical variables
the mass density $R$, two components of the momentum density $M_{\rho}$ and
$M_{z}$, and the total energy density $E$ relative to the laboratory frame of
reference.  The gas is assumed to be inviscid and compressible with an ideal
equation of state with constant adiabatic index $\Gamma$. Using cylindrical
coordinates and defining the vector
\begin{equation}
 U =  (R,M_{\rho},M_{z},E)^{T} ,
\end{equation}
the two flux vectors
\begin{equation}
F^{\rho} = (R v^{\rho},M_{\rho} v^{\rho} + p, M_{z} v^{\rho},(E +
p)v^{\rho})^{T} ,
\end{equation}
and
\begin{equation}
F^{z} = (Rv^{z},M_{\rho} v^{z},M_{z} v^{z} + p,(E + p)v^{z})^{T} ,
\end{equation}
and the source vector
\begin{equation}
S = (0,p/{\rho},0,0)^{T} ,
\end{equation}
the almost-conservative form of the equations is:
\begin{equation}
\frac{\partial{ U}}{\partial{t}} + \frac
{1}{\rho}\frac{\partial}{\partial{\rho}} (\rho F^{\rho}) +
\frac{\partial}{\partial z} (F^{z}) = S .
\end{equation}
The pressure is given by the ideal gas equation of state
\begin{equation}
p = (\Gamma - 1) (e - n) ,
\end{equation}
where $e$ and $n$ are respectively the rest frame energy density and mass
density.  In addition to their use in the equation of state they are needed
in the computation of the relativistic sound speed and in the construction of
the wave speeds.  The laboratory and rest frame variables are related via a
Lorentz transformation:
\begin{equation}
R = \gamma n ,
\end{equation}
\begin{equation}
M_{\rho} = \gamma^2 ( e + p ) v^{\rho} , \ \ \ \
M_{z} = \gamma^2 ( e + p ) v^{z} ,
\end{equation}
\begin{equation}
E = \gamma^2 ( e + p ) - p ,
\end{equation}
where $\gamma = ( 1 - v^2 )^{-1/2}$ is the Lorentz factor and $v^2 =
(v^{\rho})^2 + (v^{z})^2$.

  For this work we have chosen one of the Godunov-type solvers, well known
for their capability as robust, conservative flow solvers with excellent
shock capturing features.  In this family of solvers one reduces the problem
of updating the components of the vector $U$, averaged over a cell, to the
computation of fluxes at the cell interfaces. In one spatial dimension the
part of the update due to advection of the vector $U$ may be written as
\begin{equation}
{U^{n+1}}_{i} = {U^{n}}_{i} - \frac{\delta t}{\delta x} (F_{i + \frac{1}{2} } -
F_{i - \frac{1}{2}}) .
\end{equation}
In the scheme originally devised by Godunov (1959), a fundamental emphasis is
placed on the strategy of decomposing the problem into many local Riemann
problems, one for each pair of values of $U_{i}$ and $U_{i+1}$ to yield
values which allow the computation of the local interface fluxes
$F_{i+\frac{1}{2}}$.  In general, an initial discontinuity at $i+\frac{1}{2}$
due to $U_{i}$ and $U_{i+1}$ will evolve into four piecewise constant states
separated by three waves. The left-most and right-most waves may be either
shocks or rarefaction waves, while the middle wave is always a contact
discontinuity. The determination of these four piecewise constant states can,
in general, be achieved only by iteratively solving nonlinear equations.
Thus the computation of the fluxes necessitates a step which can be
computationally expensive.  For this reason much attention has been given to
approximate, but sufficiently accurate, techniques.  One notable method is
that due to Harten, Lax, \& Van Leer (1983; HLL), in which the middle wave,
and the two constant states that it separates, are replaced by a single
piecewise constant state.  One benefit of this approximation, which smears
the contact discontinuity somewhat, is to eliminate the iterative step, thus
significantly improving efficiency.  However, the HLL method requires
accurate estimates of the wave speeds for the left- and right-moving waves.
Einfeldt (1988) analyzed the HLL method and found good estimates for the wave
speeds. The resulting method combining the original HLL method with
Einfeldt's improvements, often referred to as the HLLE method,  has been
taken as a starting point for our simulations.

  As indicated above, one needs good wave speed estimates in using the HLLE
method. In our implementation we use estimates based on a simple application
of the relativistic addition of velocities formula for the individual
components of the velocities $v^{\rho}$ and $v^{z}$ and the relativistic
sound speed $c_s$, assuming that the waves can be decomposed into components
moving perpendicular to the two coordinate directions.

In order to compute the pressure $p$ and sound speed $c_s$ we need the rest
frame mass density $n$ and energy density $e$.  However, these quantities are
nonlinearly coupled to the components of the velocity $v^{\rho}$ and $v^{z}$,
as well as to the laboratory frame variables $R$, $M_{\rho}$, $M_{z}$, and
$E$ via the Lorentz transformation.  When the adiabatic index is constant it
is possible to reduce the computation of $n$, $e$, $v^{\rho}$, and $v^{z}$ to
the solution of the following quartic equation (cf. Schneider et al. 1993) in
the magnitude of the velocity $v$:
\begin{equation}
 \left[\Gamma v \left(E-Mv\right)-M\left(1-v^2\right)\right]^2 -\left(
 1-v^2\right)v^2\left(\Gamma-1\right)^2R^2=0 .
\end{equation}
Component velocities are then given by
\begin{equation}
v^{\rho} = {\rm sign}(M_{\rho}) v , \ \ \ \
v^{z} = M_{z} \frac{v^{\rho}}{M_{\rho}} .
\end{equation}
Then the quantities $e$ and $n$ can be found from the relations
\begin{equation}
e = E - M_{\rho} v^{\rho} - M_{z} v^{z} ,
\end{equation}
and
\begin{equation}
n = \frac{R}{\gamma} .
\end{equation}
 The quartic is solved at each cell several times during the update of a
given mesh using Newton-Raphson iteration. The quartic has an analytic
solution, but evaluation of this, and in particular, ensuring that the
correct root is chosen, proved to take longer than the simple numerical root
finder, whose accuracy improves by approximately one decimal place per
iteration cycle. Alternative root finders ({\it e.g.}, the `Brent' method)
were explored, but were slightly slower, and offered no significant advantage
in terms of robustness. It is also possible to recast equation (11) in terms
of a scalar, $u$, derived from the four-velocity, $u^i= \gamma v^i$.  This
quantity has asymptotic behavior like $v$ and $\gamma$, for slow and fast
flows respectively, and seems the natural choice of variable in the current
context. However, on exploring the form of $f(u)\equiv 0$ we found maxima and
minima such that a poor choice of starting value would have led to rapid
divergence from the desired root.

 The HLLE solver in first order mode ({\it i.e.}, with the state variables
taken as piecewise constant in each cell) is very diffusive, and produces
unacceptable results. It is therefore necessary to use, whenever possible, a
scheme of second order accuracy in both space and time. This entails in part
taking the state variables as piecewise linear in each cell.  However, in
estimating the laboratory frame values on each cell boundary, it is possible
that through discretization, the set of values $(R,M_{\rho},M_{z},E)$ is not
physical. With $M = \sqrt{M_{\rho}^2 + M_{z}^2}$, from equation (11) we see
that the requirement that $v<1$ corresponds to $M/E<1$. We also require that
$e>n$; as $e=n$ corresponds to $p=0$, we have from the Lorentz transformation
the condition $R/E < \sqrt{1-(M/E)^2}$. At each point where a transformation
is needed, we check that these conditions on $M/E$ and $R/E$ are satisfied,
and if not, recompute cell interface values in the piecewise constant
approximation.  We find that such `fall back to first order' rarely occurs.
For example, in run D (see \S 4), it occurs less than ten times per cycle,
which is a negligible fraction of the total number of transformations
performed, and is entirely restricted to a shear layer at the edge of the
grid where the jet enters the computational volume.

 We have carefully monitored the progress of each Lorentz transformation,
issuing warnings if a) a computed $v$ will lie within $\delta v$ of unity
(where $\delta v$ is set by the accuracy required of the iteration, and the
test is performed using an analytic solution obtained by expanding the
quartic about $v=1$); b) the iteration requires an excessive number of
cycles; or c) an aphysical solution ($p<0$) appears. No such problems have
been seen in any run performed to date, demonstrating the robustness of our
methods.

\section{AMR}
  The relativistic HLLE (RHLLE) method discussed in the previous section
constitutes the basic flow integration scheme on a single mesh.  The present
work also utilized adaptive mesh refinement (AMR) in order to gain spatial
and temporal resolution.

  The AMR algorithm used is a general purpose mesh refinement scheme written
by Quirk (1991). It was an outgrowth of original work by Berger (1982) and
Berger and Colella (1989). Quirk's design and implementation has proved
efficient, able to handle a wide variety of problems (see Quirk 1994),
and flexible enough to allow a variety of flow solvers to be inserted with
only modest effort.  The AMR method uses a hierarchical collection of grids
consisting of embedded meshes to discretize the flow domain.  We have used a
scheme which subdivides the domain into logically rectangular meshes with
uniform spacing in the $\rho$ and $z$ directions in a given mesh.  The AMR
algorithm orchestrates i) the flagging of cells which need further
refinement, assembling collections of such cells into meshes; ii) the
construction of boundary zones so that a given mesh is a self-contained
entity consisting of the interior cells and the needed boundary information;
iii) mechanisms for sweeping over all levels of refinement and over each mesh
in a given level to update the physical variables on each such mesh; and iv)
the transfer of data between various meshes in the hierarchy, with the
eventual completed update of all variables on all meshes to the same final
time level.  Further, the adaption of the meshes is performed so that the
data are communicated among the levels of refined meshes to ensure that the
conservative nature of the flow is preserved. The adaption process is dynamic
so that the AMR algorithm places further resolution where and when it is
needed, as well as removing resolution when it is no longer required.
Adaption occurs in time, as well as in space: the time step on a refined
grid is less than that on the coarser grid, by the larger of the refinement
factors for the two spatial dimensions. More time steps are taken on finer
grids, and the advance of the flow solution is synchronized by interleaving
the integrations at different levels. This helps prevent any interlevel
mismatches that could adversely affect the accuracy of the simulation.

  Since the tracking of shock waves is of paramount importance, effort has
been made to design features into the AMR algorithm which will prevent loss
of accuracy for shocks which extend over more than one level of refinement. A
facility for flagging a few extra cells near the shock, so that the refined
meshes will wholly contain the shock, prevents the leaking of the shock and
subsequent loss of accuracy.

  In order for the AMR method to sense where further refinement is needed,
some monitoring function is required.  We have used the gradient of the
laboratory frame mass density.  Where the gradient magnitude exceeds some
tolerance, the cell is flagged for refinement.  While other monitor functions
could be used, we have found that the density gradient is adequate for the
flows we have simulated, from non-relativistic to ultra-relativistic jets.
The success of a monitor function is possibly somewhat problem-dependent, and
merits exploration.

  The present simulations were performed using one level of refinement by a
factor of four for both directions.  During a given simulation, the AMR
algorithm managed the adaption process automatically, with no input from the
user.  The combined effect of using the RHLLE single mesh solver and the AMR
algorithm resulted in a very efficient scheme.  Where the RHLLE method was
unable to give adequate resolution on a single coarse mesh the AMR algorithm
placed more cells, resulting in an excellent overall coverage of the
computational domain.

\section{Results}

 We present the results of one nonrelativistic run (A: $v=0.3$,
$\Gamma=5/3$), and three relativistic runs (B: $\gamma=5$, $\Gamma=5/3$; C:
$\gamma=10$, $\Gamma=5/3$; and D: $\gamma=10$, $\Gamma=4/3$).  In all cases
the computational volume was 250 cells long on the coarsest grid, and the jet
had a radius of 6 cells. In runs A and B the extent and strength of the bow
shock allowed us to restrict the lateral extent of the grid to 60 cells (10
jet radii); in the higher $\gamma$ cases, the grid was extended to 100 cells
to minimize the influence of the upper boundary. The results that we present
display only the lower 60\% of the grid in these cases.  All velocities were
initially zero, except for the axial component within the 6 `inflow' cells on
the left boundary. Mass and energy densities were set by assigning the jet a
rest frame mass density of 1.0, and specifying a Mach number (${\cal M} =
\gamma\beta/\gamma_s\beta_s$) for the flow: 6 for run A, 8 for runs B and C,
15 for run D. The need to change Mach number between runs C and D arises from
the need to restrict the sound speed to lie below the maximum value permitted
for a given $\Gamma$. The speed and Mach number fix the jet sound speed, and
hence the energy density (or pressure) for a given mass density. The ambient
density was taken to be ten times that of the rest frame jet density, and the
jet and ambient pressures were taken to be equal. The ambient energy density
was then fixed.

The panels of Fig. 1 are schlieren-type images, in which the {\it gradient}
of the laboratory frame density is rendered as a grey-scale using an
exponential function; this facilitates reproduction of both the strong
features that eventually leave the computing volume, and weaker, quasi-steady
jet structures that are of more interest. A strong bow shock and contact
surface are evident in all simulations. The shape of the bow is determined in
part by the shape and lateral size of the tip of the jet -- which behaves, in
a first approximation, like a blunt obstacle; the higher Mach number of run D
causes a discernible change in global morphology in that case.  A pattern of
incident and reflection shocks and a Mach disk are evident in the
non-relativistic case (Fig. 1a), while they are far less pronounced in the
ultrarelativistic run (Fig. 1d). The wavelength of these features will be
approximately $2{\cal M}_j r_j$, the distance a flow element travels during
reflection of a sound wave from the axis. This is indeed what we find for run
A. The higher Mach number of run D implies a correspondingly longer
wavelength; there, internal structure is less sharply defined, but to the
extent that a wavelength can be measured, it agrees with that predicted.
Associated with the internal jet structure of run D, we find substantial
variations of the Lorentz factor -- which will have a profound effect on the
emergent radiation. Although the jet structure becomes more evident once the
strong bow shock has left the grid, reducing the dynamic range necessary to
render all structures, we evidently need more resolution in this part of the
flow in order to define better its interesting behavior.

A Kelvin-Helmholtz instability at the contact surface develops rapidly in the
non-relativistic case, less so for $\gamma=5$, and is not evident for the
fastest jet. Note that this is instability that develops `naturally', due,
for example, to perturbations arising from unsteadiness of the jet as
internal structure develops -- it is not due to an applied perturbation.  It
is evident from the figure that for a jet that has propagated a certain
number of jet radii, instability is much less pronounced if the flow is
highly relativistic.  A run similar to that of run B, but with $\Gamma=4/3$,
shows that a reduction in adiabatic index renders the flow somewhat more
stable, but does not prevent instability developing. Note that there is no
change in Mach number between runs B and C, and thus that the enhanced
stability is associated with the change in Lorentz factor. We have followed
the development of the flow for both cases A and D, and find that no
discernible instability arises in the latter case, during a time in which the
case D jet propagates much further than the case A jet -- many hundred jet
radii, although we have not as yet been able to pursue case D for the total
time taken on run A. A boundary `reflection' occurs when a strong bow shock
nearly normal to the boundary crosses the edge of the computing volume (see
Fig. 1b). Removal of this effect at outflow boundaries would require modeling
the bow `wave' within the ghost zones, and even then, some residual
reflection can be expected.  Furthermore, the propagation off the grid of
strong features, such as the contact surface at the head of the flow, can
lead to disturbances that propagate back into the flow (Hardee, private
communication).  We were, therefore, cautious when interpreting runs in which
the dominant structures have left the computational volume, but do not
believe that such effects play a significant role in influencing flow
stability.

Run C exhibits a shock in the previously-shocked ambient medium, some hint of
which may be seen in runs B and D. The upstream Mach number of this feature
is $\sim 2$ near to the jet, approaching $\sim 1$ as one moves towards the
upper left. Such features are well-known in non-relativistic flows past blunt
obstacles ({\it e.g.}, Zucrow \& Hoffman 1985), and arise from the rapid
expansion, with associated acceleration and cooling, of the hot, shocked
material. We note that the evidence of this feature is a function of flow
speed and adiabatic index.

\section{Discussion}

We have described a numerical hydrodynamical code, and the first results
there from, which provides a robust and efficient method for simulating
relativistic jets. The above examples were run on IBM R6000, Sun Sparc 10 and
SGI Indigo and Onyx  workstations. It is thus possible, using widely
available computing resources, to produce detailed quantitative models of the
dynamics of parsec scale flows, although 3-D computations will demand
supercomputer facilities.

The primary conclusion of this study is that jet stability is strongly
influenced by Lorentz factor: we see a dramatic reduction in the development
of the Kelvin-Helmholtz mode as the Lorentz factor is increased, with no
evidence of instability for $\gamma\sim 10$.  There is now convincing
evidence that the flow speeds of BL~Lacs are systematically less than that of
QSOs, while they show evidence of a more turbulent internal structure and
discrete knots, in contrast to the ordered field and smoother intensity
distribution of QSOs (Gabuzda et al. 1989; Gabuzda et al. 1994). Our
simulations suggest that this might be explained by a transition to a more
stable condition for the higher Lorentz factor QSOs.

We intend to refine the solver, and explore more general equations of state.
However, we see the most important development of this work as being the
induction of perturbations, leading to propagating structures within the
jet, and the performance of detailed radiation transfer calculations through
these flows for comparison with VLBI and VLBP images.

\acknowledgments

This work was supported by NSF grant AST 9120224 and by the Ohio
Supercomputer Center from a Cray Research Software Development Grant.  We
wish to thank James Quirk for providing the source code for AMR, and much
advice on its use. We are very grateful to David Garfinkle for helpful
discussions, Douglas Richstone for the loan of an IBM R6000, and the Ohio
Visualization Laboratory for providing access to a SGI Onyx for a portion of
the computations.

\clearpage

\begin{figure}
\caption{Schlieren-type images of laboratory frame density: a. $v=0.3\ \Gamma
=5/3$, b. $\gamma=5\ \Gamma=5/3$, c. $\gamma=10\ \Gamma=5/3$, d. $\gamma=10 \
\Gamma=4/3$. Note that such schlieren-type images accentuate weak features,
and that the `staircasing' in panel c. is an artifact of the plotting.}
\end{figure}

\end{document}